\documentclass[10pt,a4paper,onecolumn]{article}
\usepackage{marginnote}
\usepackage{graphicx}
\usepackage{xcolor}
\usepackage{authblk,etoolbox}
\usepackage{titlesec}
\usepackage{calc}
\usepackage{tikz}
\usepackage{hyperref}
\hypersetup{colorlinks,breaklinks=true,
            urlcolor=[rgb]{0.0, 0.5, 1.0},
            linkcolor=[rgb]{0.0, 0.5, 1.0}}
\usepackage{caption}
\usepackage{tcolorbox}
\usepackage{amssymb,amsmath}
\usepackage{ifxetex,ifluatex}
\usepackage{seqsplit}
\usepackage{xstring}

\usepackage{float}
\let\origfigure\figure
\let\endorigfigure\endfigure
\renewenvironment{figure}[1][2] {
    \expandafter\origfigure\expandafter[H]
} {
    \endorigfigure
}

\usepackage{fixltx2e} 
\usepackage[
  backend=biber,
]{biblatex}
\bibliography{paper.bib}

\let\textttOrig=\texttt
\def\texttt#1{\expandafter\textttOrig{\seqsplit{#1}}}
\renewcommand{\seqinsert}{\ifmmode
  \allowbreak
  \else\penalty6000\hspace{0pt plus 0.02em}\fi}

\makeatletter
\let\href@Orig=\href
\def\href@Urllike#1#2{\href@Orig{#1}{\begingroup
    \def\Url@String{#2}\Url@FormatString
    \endgroup}}
\def\href@Notdoi#1#2{\def\tempa{#1}\def\tempb{#2}%
  \ifx\tempa\tempb\relax\href@Urllike{#1}{#2}\else
  \href@Orig{#1}{#2}\fi}
\def\href#1#2{%
  \IfBeginWith{#1}{https://doi.org}%
  {\href@Urllike{#1}{#2}}{\href@Notdoi{#1}{#2}}}
\makeatother

\newlength{\cslhangindent}
\newlength{\csllabelwidth}
\newenvironment{CSLReferences}[3] 
 {
  \setlength{\parindent}{0pt}
  \ifodd #1 \everypar{\setlength{\hangindent}{\cslhangindent}}\ignorespaces\fi
  \ifnum #2 > 0
  \setlength{\parskip}{#2\baselineskip}
  \fi
 }%
 {}
\usepackage{calc}

\usepackage[top=3.5cm, bottom=3cm, right=1.5cm, left=1.0cm,
            headheight=2.2cm, reversemp, includemp, marginparwidth=4.5cm]{geometry}

\titleformat{\section}
  {\normalfont\sffamily\Large\bfseries}
  {}{0pt}{}
\titleformat{\subsection}
  {\normalfont\sffamily\large\bfseries}
  {}{0pt}{}
\titleformat{\subsubsection}
  {\normalfont\sffamily\bfseries}
  {}{0pt}{}
\titleformat*{\paragraph}
  {\sffamily\normalsize}

\usepackage{fancyhdr}
\makeatletter
\let\ps@plain\ps@fancy
\fancyheadoffset[L]{4.5cm}
\fancyfootoffset[L]{4.5cm}

\definecolor{linky}{rgb}{0.0, 0.5, 1.0}

\newtcolorbox{repobox}
   {colback=red, colframe=red!75!black,
     boxrule=0.5pt, arc=2pt, left=6pt, right=6pt, top=3pt, bottom=3pt}

\newcommand{\ExternalLink}{%
   \tikz[x=1.2ex, y=1.2ex, baseline=-0.05ex]{%
       \begin{scope}[x=1ex, y=1ex]
           \clip (-0.1,-0.1)
               --++ (-0, 1.2)
               --++ (0.6, 0)
               --++ (0, -0.6)
               --++ (0.6, 0)
               --++ (0, -1);
           \path[draw,
               line width = 0.5,
               rounded corners=0.5]
               (0,0) rectangle (1,1);
       \end{scope}
       \path[draw, line width = 0.5] (0.5, 0.5)
           -- (1, 1);
       \path[draw, line width = 0.5] (0.6, 1)
           -- (1, 1) -- (1, 0.6);
       }
   }

\patchcmd{\@maketitle}{center}{flushleft}{}{}
\patchcmd{\@maketitle}{center}{flushleft}{}{}
\patchcmd{\@maketitle}{\LARGE}{\LARGE\sffamily}{}{}
\def\maketitle{{%
  
  \AB@maketitle}}
\makeatletter
\renewcommand\AB@affilsepx{ \protect\Affilfont}
\renewcommand\AB@affilnote[1]{{\bfseries #1}\hspace{3pt}}
\renewcommand{\affil}[2][]%
   {\newaffiltrue\let\AB@blk@and\AB@pand
      \if\relax#1\relax\def\AB@note{\AB@thenote}\else\def\AB@note{#1}%
        \setcounter{Maxaffil}{0}\fi
        \begingroup
        \let\href=\href@Orig
        \let\texttt=\textttOrig
        \let\protect\@unexpandable@protect
        \def\thanks{\protect\thanks}\def\footnote{\protect\footnote}%
        \@temptokena=\expandafter{\AB@authors}%
        {\def\\{\protect\\\protect\Affilfont}\xdef\AB@temp{#2}}%
         \xdef\AB@authors{\the\@temptokena\AB@las\AB@au@str
         \protect\\[\affilsep]\protect\Affilfont\AB@temp}%
         \gdef\AB@las{}\gdef\AB@au@str{}%
        {\def\\{, \ignorespaces}\xdef\AB@temp{#2}}%
        \@temptokena=\expandafter{\AB@affillist}%
        \xdef\AB@affillist{\the\@temptokena \AB@affilsep
          \AB@affilnote{\AB@note}\protect\Affilfont\AB@temp}%
      \endgroup
       \let\AB@affilsep\AB@affilsepx
}
\makeatother

\renewcommand\Affilfont{\sffamily\small\mdseries}
\ifnum 0\ifxetex 1\fi\ifluatex 1\fi=0 
  \usepackage[T1]{fontenc}
  \usepackage[utf8]{inputenc}

\else 
  \ifxetex
    \usepackage{mathspec}
    \usepackage{fontspec}

  \else
    \usepackage{fontspec}
  \fi
  \defaultfontfeatures{Ligatures=TeX,Scale=MatchLowercase}

\fi
\IfFileExists{upquote.sty}{\usepackage{upquote}}{}
\IfFileExists{microtype.sty}{%
\usepackage{microtype}
\UseMicrotypeSet[protrusion]{basicmath} 
}{}

\usepackage{hyperref}
\hypersetup{unicode=true,
            pdftitle={Siril: An Advanced Tool for Astronomical Image Processing},
            pdfborder={0 0 0},
            breaklinks=true}
\let\addcontentslineOrig=\addcontentsline
\def\addcontentsline#1#2#3{\bgroup
  \let\texttt=\textttOrig\addcontentslineOrig{#1}{#2}{#3}\egroup}
\let\markbothOrig\markboth
\def\markboth#1#2{\bgroup
  \let\texttt=\textttOrig\markbothOrig{#1}{#2}\egroup}
\let\markrightOrig\markright
\def\markright#1{\bgroup
  \let\texttt=\textttOrig\markrightOrig{#1}\egroup}

\usepackage{graphicx,grffile}
\makeatletter
\def\maxwidth{\ifdim\Gin@nat@width>\linewidth\linewidth\else\Gin@nat@width\fi}
\def\maxheight{\ifdim\Gin@nat@height>\textheight\textheight\else\Gin@nat@height\fi}
\makeatother
\setkeys{Gin}{width=\maxwidth,height=\maxheight,keepaspectratio}
\IfFileExists{parskip.sty}{%
\usepackage{parskip}
}{
\setlength{\parindent}{0pt}
\setlength{\parskip}{6pt plus 2pt minus 1pt}
}
\providecommand{\tightlist}{%
  \setlength{\itemsep}{0pt}\setlength{\parskip}{0pt}}
\ifx\paragraph\undefined\else
\let\oldparagraph\paragraph
\renewcommand{\paragraph}[1]{\oldparagraph{#1}\mbox{}}
\fi
\ifx\subparagraph\undefined\else
\let\oldsubparagraph\subparagraph
\renewcommand{\subparagraph}[1]{\oldsubparagraph{#1}\mbox{}}
\fi

\title{Siril: An Advanced Tool for Astronomical Image Processing}

        \author[1]{Cyril Richard}
          \author[2]{Vincent Hourdin}
          \author[2]{Cécile Melis}
          \author[3]{Adrian Knagg-Baugh}
    
      \affil[1]{Laboratoire Interdisciplinaire Carnot de Bourgogne, UMR
6303 CNRS - Université de Bourgogne, 9 Av. A. Savary, BP 47870, F-21078
Dijon Cedex, France}
      \affil[2]{Independent Researcher, France}
      \affil[3]{Independent Researcher, United Kingdom}
  \date{\vspace{-7ex}}

\begin{document}
\maketitle

\marginpar{

  \begin{flushleft}
  \sffamily\small

  {\bfseries DOI:} \href{https://doi.org/DOI unavailable}{\color{linky}{DOI unavailable}}

  \vspace{2mm}

  {\bfseries Software}
  \begin{itemize}
    \setlength\itemsep{0em}
    \item \href{N/A}{\color{linky}{Review}} \ExternalLink
    \item \href{NO_REPOSITORY}{\color{linky}{Repository}} \ExternalLink
    \item \href{DOI unavailable}{\color{linky}{Archive}} \ExternalLink
  \end{itemize}

  \vspace{2mm}

  \par\noindent\hrulefill\par

  \vspace{2mm}

  {\bfseries Editor:} \href{https://example.com}{Pending
Editor} \ExternalLink \\
  \vspace{1mm}
    {\bfseries Reviewers:}
  \begin{itemize}
  \setlength\itemsep{0em}
    \item \href{https://github.com/Pending Reviewers}{@Pending
Reviewers}
    \end{itemize}
    \vspace{2mm}

  {\bfseries Submitted:} N/A\\
  {\bfseries Published:} N/A

  \vspace{2mm}
  {\bfseries License}\\
  Authors of papers retain copyright and release the work under a Creative Commons Attribution 4.0 International License (\href{http://creativecommons.org/licenses/by/4.0/}{\color{linky}{CC BY 4.0}}).

  \end{flushleft}
}

\hypertarget{summary}{%
\section{Summary}\label{summary}}

Siril is a powerful open source software package designed for the pre-
and post-processing of astronomical images. It is well suited for
astrophotography enthusiasts and professional astronomers alike. Siril
provides advanced tools for tasks such as image stacking, calibration,
registration, and enhancement, enabling users to produce high quality
images of celestial objects.

\hypertarget{statement-of-need}{%
\section{Statement of Need}\label{statement-of-need}}

Astronomical imaging requires specialized software capable of handling
the unique challenges presented by the data. Siril addresses these
challenges by providing a suite of tools designed to process images from
various types of astronomical instruments. Its robust feature set
includes support for multiple image formats and precise photometric
calibration techniques.

Siril stands out for its user-friendly interface and integration with
other astronomical software packages. It offers a comprehensive solution
for both amateur and professional astronomers to enhance their imaging
workflows. The software has been utilized in numerous scientific
publications and astrophotography projects, demonstrating its
versatility and effectiveness.

There are many solutions available for astronomical imaging, but few are
open source. For example, IRIS was a popular tool but has not been
developed since 2005. Another well-known software, IRAF, consists of
40-year-old legacy code, and institutional support for IRAF and its use
is rapidly declining. It is recommended to search for alternative
solutions, such as those in the Astropy community, and to avoid starting
new projects using IRAF (Ogaz \& Tollerud, 2018). Siril fills this gap
by providing a modern, open source alternative that continues to evolve
and support the needs of the astronomical community.

\hypertarget{features-and-functionality}{%
\section{Features and Functionality}\label{features-and-functionality}}

Siril provides a range of features to support astronomical image
processing:

\begin{itemize}
\item
  \textbf{Image calibration}: Correction of biases, darks, and flats to
  calibrate astronomical images data.
\item
  \textbf{Image registration and stacking}: Alignment of images taken at
  different times and their subsequent stacking to increase
  signal-to-noise ratio. This version also incorporates the Hubble Space
  Telescope drizzle algorithm (Fruchter \& Hook, 2002) for applying WCS
  (World Coordinate System) and registration data transforms, providing
  improved detail reconstruction when processing sets of under-sampled
  images.
\item
  \textbf{Advanced image enhancement}: Application of various filters
  and algorithms to improve image details and reduce noise. For example,
  the Larson-Sekanina filter is particularly useful for highlighting
  non-circular structures in images of comets by enhancing radial
  features and making structures like jets and dust trails more visible
  (Larson \& Sekanina, 1984). An example of such a filter applied to a
  comet image is given in \autoref{fig:rgradient}.

  \begin{figure}
  \centering
  \includegraphics{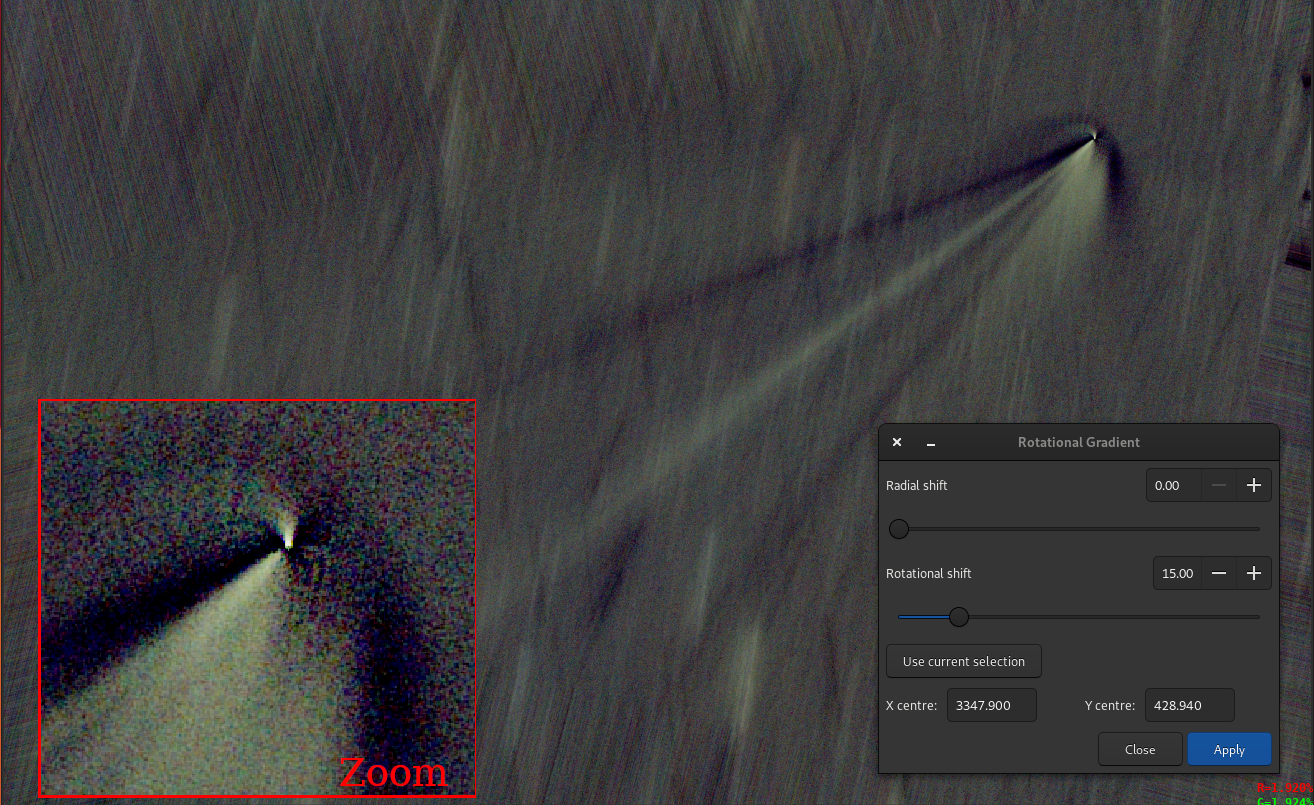}
  \caption{Application of the Larson-Sekanina filter on a comet image to
  highlight non-circular structures in the image.\label{fig:rgradient}}
  \end{figure}
\item
  \textbf{Astrometry}: Precise measurement and analysis of the positions
  and movements of celestial objects within the images. Siril includes
  functionalities for astrometric calibration, allowing users to find an
  astrometric solution and match observed star and solar system object
  positions with reference catalogs. It uses the WCS (World Coordinate
  System) features from FITS file keywords (Calabretta et al., 2004;
  Calabretta \& Greisen, 2002; Greisen \& Calabretta, 2002).
\item
  \textbf{Spectro photometric calibration}: Accurate calibration of
  image photometry using the latest version of the Gaia catalog (Gaia
  Collaboration et al., 2023) for recovering the true colors of
  astronomical objects in the images.
\item
  \textbf{Photometry}: Capability to perform photometric analysis,
  including the creation of light curves for variable stars and the
  observation of exoplanet transits.
\item
  \textbf{Scriptability}: Ability to automate repetitive tasks through
  scripting, increasing efficiency.
\end{itemize}

These are just a selection of the many features Siril offers. The number
of functionalities is significantly larger and continuously growing as
the software evolves.

\hypertarget{freeastro-ecosystem}{%
\section{FreeAstro Ecosystem}\label{freeastro-ecosystem}}

The FreeAstro ecosystem encompasses various projects and repositories
that support and extend the functionality of Siril. FreeAstro,
\href{https://gitlab.com/free-astro}{hosted on GitLab}, serves as the
umbrella organization for all projects related to Siril, fostering a
collaborative environment for development, documentation, and community
engagement.

\begin{itemize}
\tightlist
\item
  \href{https://gitlab.com/free-astro/siril-web}{siril-web}: The
  \href{https://siril.org}{official website} for Siril, built using the
  open source static site generator Hugo and hosted by
  \href{https://pixls.us}{pixls.us}, which also hosts forums for open
  source image processing software. This site provides users with access
  to the latest news, updates, and resources related to Siril.
\item
  \href{https://gitlab.com/free-astro/siril-doc}{siril-doc}: The
  documentation repository, hosted by ReadTheDocs. This repository
  contains comprehensive user manuals, guides, and reference materials
  to assist users in effectively utilizing Siril's features. There are
  two branches available: one for the
  \href{https://siril.readthedocs.io/en/stable}{stable version} and one
  for the \href{https://siril.readthedocs.io/en/latest}{development
  version}.
\item
  \href{https://gitlab.com/free-astro/siril-localized-doc}{siril-localized-doc}:
  Dedicated to the translation of Siril's documentation. This repository
  leverages Weblate, a translation platform, to support multiple
  languages and ensure that documentation is accessible to a global
  audience.
\item
  \href{https://gitlab.com/free-astro/siril-spcc-database}{siril-spcc-database}:
  This repository stores data related to the Spectro Photometric Color
  Calibration (SPCC) tool. It includes JSON files detailing
  OSC/monochrome sensors and filters available on the market. The
  primary goal of this database is to collect extensive data and promote
  collaboration within the astronomy community.
\item
  \href{https://gitlab.com/free-astro/siril-scripts}{siril-scripts}: A
  repository where users can share and contribute scripts that extend
  the functionality of Siril. These user-contributed scripts are
  integrated into Siril, providing additional tools and capabilities for
  the community.
\end{itemize}

\hypertarget{acknowledgements}{%
\section{Acknowledgements}\label{acknowledgements}}

We acknowledge contributions from the Siril development community.
Special thanks to our testers and users who provided invaluable
feedback. We also extend our gratitude to François Meyer, who initiated
the project over 20 years ago. Additionally, we thank the team at
pixls.us for generously hosting our website and forums, supporting the
open source image processing community.

\hypertarget{references}{%
\section*{References}\label{references}}
\addcontentsline{toc}{section}{References}

\hypertarget{refs}{}
\begin{CSLReferences}{1}{0}
\leavevmode\hypertarget{ref-wcsII}{}%
Calabretta, M. R., \& Greisen, E. W. (2002). Representations of
celestial coordinates in FITS. \emph{Astronomy \& Astrophysics},
\emph{395}(3), 1077--1122.
\url{https://doi.org/10.1051/0004-6361:20021327}

\leavevmode\hypertarget{ref-wcsIV}{}%
Calabretta, M. R., Valdes, F., Greisen, E. W., \& Allen, S. L. (2004).
Representations of distortions in FITS world coordinate systems.
\emph{Astronomical Data Analysis Software and Systems (ADASS) XIII},
\emph{314}, 551.

\leavevmode\hypertarget{ref-fruchter2002drizzle}{}%
Fruchter, A., \& Hook, R. (2002). Drizzle: A method for the linear
reconstruction of undersampled images. \emph{Publications of the
Astronomical Society of the Pacific}, \emph{114}(792), 144.
\url{https://doi.org/10.1086/338393}

\leavevmode\hypertarget{ref-refgaia}{}%
Gaia Collaboration, Vallenari, A., Brown, A. G. A., Prusti, T., de
Bruijne, J. H. J., Arenou, F., Babusiaux, C., Biermann, M., Creevey, O.
L., Ducourant, C., Evans, D. W., Eyer, L., Guerra, R., Hutton, A.,
Jordi, C., Klioner, S. A., Lammers, U. L., Lindegren, L., Luri, X.,
\ldots{} Zwitter, T. (2023). {Gaia Data Release 3 - Summary of the
content and survey properties}. \emph{Astronomy \& Astrophysics},
\emph{674}, A1. \url{https://doi.org/10.1051/0004-6361/202243940}

\leavevmode\hypertarget{ref-wcsI}{}%
Greisen, E. W., \& Calabretta, M. R. (2002). Representations of world
coordinates in FITS. \emph{Astronomy \& Astrophysics}, \emph{395}(3),
1061--1075. \url{https://doi.org/10.1051/0004-6361:20021326}

\leavevmode\hypertarget{ref-larson1984coma}{}%
Larson, S., \& Sekanina, Z. (1984). {Coma morphology and dust-emission
pattern of periodic Comet Halley. I-High-resolution images taken at
Mount Wilson in 1910}. \emph{Astronomical Journal (ISSN 0004-6256), Vol.
89, April 1984, p. 571-578.}, \emph{89}, 571--578.
\url{https://doi.org/10.1086/113551}

\leavevmode\hypertarget{ref-endiras}{}%
Ogaz, S., \& Tollerud, E. (2018). Removing the institute's dependence on
{IRAF} ({Y}ou can do it too!). \emph{Space Telescope Science Institute
Newsletter}, \emph{35}(3).
\url{https://www.stsci.edu/contents/newsletters/2018-volume-35-issue-03/removing-the-institutes-dependence-on-iraf-you-can-do-it-too}

\end{CSLReferences}

\end{document}